\documentstyle[12pt,epsf]{article}
\def\a{\alpha}

\begin{document} 
\begin{titlepage}
\title{
\hfill\parbox{7cm}
{\normalsize TIFR/TH/99/52, IMSc -99/10/35}\\
Three-Manifold Invariants from \\
Chern-Simons Field Theory with Arbitrary Semi-Simple Gauge Groups}
\author{ Romesh K. Kaul$^a$ 
\footnote{E-mail: kaul@imsc.ernet.in}
and P. Ramadevi$^b$  
\footnote{E-mail: rama@theory.tifr.res.in, ramadevi@phy.iitb.ernet.in}
\\
\\
$^a$ The Institute of Mathematical Sciences, 
 Chennai 600113, India.
\\
$^b$ Physics Department, 
Indian Institute of Technology, Bombay\\
Mumbai 400 076, India.}
\date{ June 2000}

\maketitle
\thispagestyle{empty}
Invariants  for framed links in $S^3$ obtained  from 
Chern-Simons gauge field theory  based on an arbitrary gauge group 
(semi-simple) have been used to construct a three-manifold invariant. 
This is a 
generalization of a similar construction developed earlier for 
$SU(2)$ Chern-Simons
theory. The procedure exploits  a theorem of  Lickorish and Wallace  and also
those of Kirby, Fenn and Rourke  which relate three-manifolds to 
surgeries on framed unoriented links. The invariant is an appropriate
linear combination of framed link invariants which does not change under Kirby
calculus. This combination does not see the 
relative orientation of the component knots. 
The invariant is related to the partition function of
Chern-Simons theory. This thus provides an efficient method of 
evaluating the partition function for these  field theories. 
As some  examples, explicit computations 
of these manifold invariants for a few three-manifolds have been done.

\end{titlepage}

\section{Introduction}

In recent times topological quantum theories have proved to be a very powerful 
tool for the study of geometry and topology of low dimensional manifolds.
An example of such theories is the Chern-Simons  gauge field theory which provides
a general framework for knots and links in three dimensions \cite {ati}.
Vacuum expectation values of Wilson link operators in this theory
yield a class of polynomial link invariants.
It was E. Witten who in his pioneering paper about ten years ago developed
this framework and also demonstrated that the famous Jones polynomial was
related to the expectation value of a Wilson loop operator (in spin $1/2$
representation) in an $SU(2)$
Chern-Simons field theory\cite{wit}. Since then many attempts  have been made to
obtain exact and explicit non-perturbative solutions to such field theories \cite{kau1,
kau2, lab1,kau3,lab2, kau4}.
Powerful methods for completely analytical and non-perturbative computations of
the expectation values of Wilson link operators have been developed.
One such method in its complete manifestation has been presented 
in ref.\cite{kau2}. 
 
The power of field theoretic  framework  through Chern-Simons theories
is indeed so deep that it allows us to study knots and links  not only 
in simple manifolds  such as a three-sphere but also in any arbitrary
three-manifold. For example, the link invariants obtained in these
field theories can be used to construct three-manifold invariants.
One such construction involves an application of Lickorish-Wallace 
surgery presentation of three-manifolds in terms of unoriented framed links
embedded in $S^3$.  Surgery on more than one framed knot or link can yield the 
same manifold.  However, the rules of equivalence of framed links which yield 
the same three-manifold on surgery are given by theorems of Kirby,
Fenn and Rourke and are  known as Kirby moves. Thus a three-manifold
can be characterized by an appropriate combination of invariants of the  
associated framed knots and links which is unchanged under Kirby moves.
A three-manifold invariant of this type has been recently constructed from
invariants for framed links in  an $SU(2)$ Chern-Simons
theory  in  a three-sphere \cite{kau3, kau4}. The algebraic formula for the
invariant so obtained is rather easy to compute for an arbitrary
three-manifold. The construction developed is general enough to
yield  other three-manifold 
invariants from the link invariants of  Chern-Simons gauge theories
based on other semi-simple gauge groups. This extension is what will 
be presented in the present paper.

Other three-manifold invariants  have  also been constructed
in recent years. For example, exploiting the surgery presentations of
three-manifolds
in terms of unoriented framed links, Lickorish had earlier obtained a manifold
invariant
using bracket polynomials of cables \cite {lic}. Evaluation of this invariant
involves a tedious calculation through recursion relations.  Using representation
theory of composite braids \cite {ram1}, it has been possible to directly evaluate
the bracket polynomials for cables without going through the recursion relations. 
This direct calculation has been used to demonstrate the equivalence
of the invariant obtained from $SU(2)$ Chern-Simons theory in ref.\cite{kau3, kau4}
to the Lickorish's three-manifold invariant up to a variable redefinition
\cite{ram2}. Further  Lickorish's  invariant  is considered to be a  reformulation
of Reshetikhin-Turaev invariant\cite {res}, which in turn is known to be equivalent
to the partition function of  $SU(2)$ Chern-Simons theory, known as Witten
invariant. Thus this establishes\cite{ram2}, by an indirect method,
that the field theoretic three-manifold invariant obtained
from  link invariants in $SU(2)$ Chern-Simons
gauge theory using  theorems of Lickorish and  Wallace, Kirby, Fenn and Rourke
is  actually  partition function of $SU(2)$ Chern-Simons theory, 
a fact already
noticed for many three-manifolds in ref.\cite{kau3}. This equivalence is up to 
an over all normalization.

The link invariants in general depend on the framing convention used. The 
{\it frame} of a knot is an associated closed curve going along the 
length of knot and wrapping around it certain number of times. In the
field theoretic language, framing has to do with the regularization
prescription used to define the coincident loop correlators \cite{kau4}. In one such
framing convention known  as {\it standard framing} the self-linking number of
every knot ({\it i.e.}, linking number of the knot and its framing curve) 
is zero. This convention was used in obtaining the link invariants
in  ref. \cite{kau2}. The invariants so obtained are ambient isotopic
invariant, that is, these are unchanged under all the three Reidemeister
moves. However, in our present discussion, we 
are interested in {\it framed links} which are only regular isotopic,
that is, two framed links are equivalent if and only if they are related
by two of the Reidmeister moves (excluding the
one that changes the writhe). The framing convention for describing
such framed links is {\it vertical framing}.
Here the frame is to be just vertically
above the strands of every knot projected on to a plane. 
The link invariants in this framing exhibit  only regular isotopy
invariance. These framed link invariants are in general 
sensitive to the relative
orientations of component knots of a link. 
Reversing the orientation in a knot component changes the representation
living on the associated Wilson loop operator to its conjugate
representation.
We construct an appropriate linear combination of these invariants
for different group representations on the framed link
which is unchanged under Kirby calculus. This combination then
characterises the manifold related to the given link by 
surgery. Though the individual terms in this combination 
in general depend on the relative orientations of the knots, the 
combination does not. This is consistent with  
Lickorish-Wallace theorem for surgery presentation of three-manifolds 
which involves only {\it unoriented} links.

The plan of the paper is as follows: In the next section, 
we shall briefly discuss Chern-Simons theory based on any arbitrary
semi-simple  gauge group.
Methods of computing the expectation value of Wilson loop operators for 
framed knots and links will be outlined. These are generalizations
of the methods for $SU(2)$ Chern-Simons theory presented in 
ref. \cite{kau2, kau3}. 
In Sec.3, we shall present a theorem of Lickorish and Wallace and Kirby calculus
which are the necessary ingredients in the construction
of three-manifolds by surgery. Using the field theoretic framed link 
invariants, we derive an algebraic formula for  a three-manifold invariant.  
Sec. 4 will contain some concluding remarks.

\section{Chern-Simons field theory and link invariants} 

In this section, we shall present some of the salient
features of Chern-Simons theory on $S^3$ based on 
arbitrary semi-simple gauge group ${\cal G}$ and the
invariants of {\it framed links} embedded in $S^3$.
These framed link invariants  will be used in the 
construction of three-manifold invariants in the next section.

For a matrix valued 
connection one-form $A$ of the gauge group ${\cal G}$, the
Chern-Simons action $S$ on $S^3$ is given by
\begin{equation}
kS ~=~ {\frac k {4\pi}} \int_{S^3} tr(AdA~+~{\frac{2}{3}} A^3)~.
\label{ac}
\end{equation}
\noindent The coupling constant $k$ takes integer values.
Clearly action (\ref{ac}) does not have any metric of $S^3$ in it.
The topological operators are the metric independent Wilson loop (knot)
operators defined as
\begin{equation} 
W_R [C]~=~ tr_R Pexp \oint_C A_R  \label{wil}
\end{equation}  
\noindent for a knot $C$ carrying representation $R$; $A_R$ is the connection
field in representation $R$ of the group.  Reversing the orientation of a knot
corresponds to placing conjugate representation $R^*$ on it.

For a link $L$ made up of component knots $C_1, C_2, \ldots C_r$
carrying $R_1,
R_2,\ldots R_r$ representations respectively, we have the
Wilson link operator
defined as
\begin{equation}
W_{R_1R_2\ldots R_r} [L] \ = \ \prod_{\ell=1}^r \ W_{R_\ell}
[C_\ell] ~.\label{wil1}
\end{equation}
\noindent We are interested in the functional averages of
these link operators:
\begin{eqnarray} 
V[L; {\bf f}; R_1, R_2  \ldots R_r] \equiv V_{R_1R_2\ldots R_r}[D_L] \ &=& 
\ Z^{-1} \int_{S^3} [dA]
W_{R_1R_2\ldots R_r} [L]
e^{ikS},  \label{jon}\\
~~{\rm where}~~ Z \ = \ \int_{S^3} [dA] e^{ikS} ~, &~&\nonumber    
\end{eqnarray}
and $D_L$ denotes link diagram corresponding to framed link $[L,f]$.
Link diagrams are obtained from a regular projection of a link in a plane
with transverse double points (refered to as crossings) 
as the only self-intersections. We work in the 
vertical frame where the frame curve is vertically above the
plane of the diagram. Hence, in vertical  framing, 
 ${\bf f}=(f_1,f_2, \ldots, f_r)$ on link $L$ is a set of integers 
denoting the sum of the
crossing signs in the part of the diagram representing the components. 
These expectation values are the {\it generalized regular isotopy invariants}
of framed links. These  can be exactly evaluated using 
the method developed in ref. \cite{kau2}. The method makes use of
the following two inputs: 
\begin{enumerate}
\item
{Chern-Simons functional integral (containing Wilson lines) 
on a three-manifold  with 
$n$-punctures on its  boundary corresponds to a state in the space of
$n$-correlator conformal blocks in the corresponding
Wess-Zumino conformal field theory on that boundary \cite {wit}.}
 
\item
{Knots and Links can be obtained by closure of braids (Alexander
theorem) or equivalently  platting of braids (Birman theorem) \cite{bir}.}
\end{enumerate}

Consider a manifold $S^3$         
from which two non-intersecting three-balls are removed.
This manifold                  
has two boundaries, each an $S^2$.  We place $2n$  Wilson line-integrals 
over lines connecting these two boundaries  through a weaving 
pattern $\bf B$ as  shown in the Figure (a) below. 
This is a $2n-$braid placed in this manifold.  
The strands are specified  on the upper boundary by giving $2n$ 
assignments $(\hat R_1^*, \hat R_1, \hat R_2^*, \hat R_2, \ldots , \hat R_n^*,
\hat R_n)$. Here $\hat R = (R, \epsilon)$ denotes representation $R$
and orientation $\epsilon$ ($\epsilon = \pm 1$ for a strand going into
or away from the boundary) and conjugate assignment 
$\hat R^* = (\bar R, - \epsilon)$ indicates reversal of the orientation.
Similar specifications are done with respect to the lower boundary by
the representation assignments $(\hat \ell_1, \hat \ell_1^*,~
\hat \ell_2,  \hat \ell_2^*, .... \hat \ell_n, \hat \ell_n^*)$.
Then the assignments $\{ \hat \ell_i \}$ are just a permutation of
$\{ \hat R_i^* \}$. Chern-Simons functional integral over this manifold 
is a state in the tensor product of the Hilbert spaces
associated with the two boundaries, ${\cal{H}}_1 \, \otimes
{\cal{H}}_2$. This state can be expanded in terms of some convenient
basis\cite{kau2, kau3}.  These bases are given by the conformal blocks for 
$2n$-point correlators of the  associated  ${\cal G}_k$ Wess-Zumino conformal 
field theory on each of the  $S^2$ boundaries. 
\begin{figure}
\centerline{\epsfxsize=5in \epsfbox {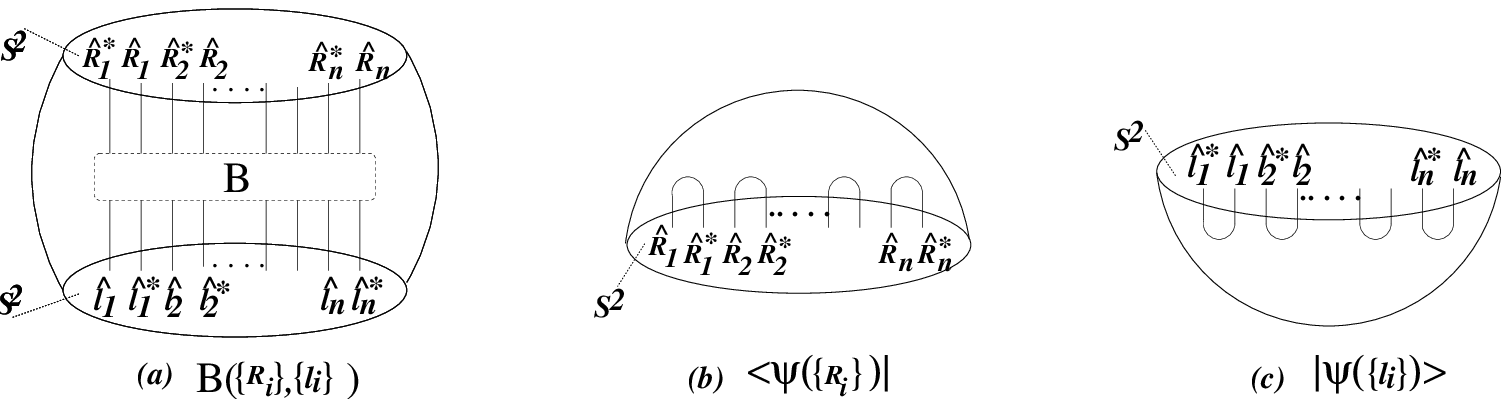}}
\end{figure} 

An arbitrary braid can be generated by a sequence of
elementary braidings. The eigenvalues  of these elementary
braids are given by conformal field theory.
The braiding eigenvalues depend on the 
framing. In vertical framing, the eigenvalues  
for a right-handed half-twist between 
two parallely oriented strands carrying representation $R,~R'$ 
($\epsilon \epsilon' = 1$) and for anti-parallely oriented strands
($\epsilon \epsilon' = -1$), as shown in the figure below, 
are respectively \cite{kau2} :
\begin{eqnarray}
\lambda^{(+)}_t (R,R')&=&  
(-)^{\epsilon_R+ \epsilon_{R'} - 
\epsilon_t}~ q^{-(C_R+C_{R'})/2  + 
C_t/2}  \\
 \lambda^{(-)}_t (R,R') &=&  
(-)^{\vert \epsilon_R- \epsilon_{R'} \vert - 
\epsilon_t}~  q^{(C_R+C_{R'})/2 - C_t/2}~, 
\end{eqnarray}
\noindent where $t$ takes values allowed in the 
product of representations of $R$ and $R'$ 
given by the fusion rules of ${\cal G}_k$
Wess-Zumino conformal field theory, $q$-independent phases 
$ (-)^{\epsilon_R + \epsilon_{R'} - \epsilon_t}=\pm 1$,
$ (-)^{\vert \epsilon_R - \epsilon_{R'} \vert - \epsilon_t}=\pm 1$,
and $C_R, C_{R'}$ 
are the quadratic Casimir of  representations $R,R'$ respectively.
The variable $q$ in the above equation is related to the coupling 
constant $k$ in Chern-Simons theory as
\begin{equation}
q= exp( \frac {2\pi i} { k+C_v})~, \label {quon}
\end{equation}
where $C_v$ is quadratic Casimir of  adjoint representation.
$$ \epsfbox{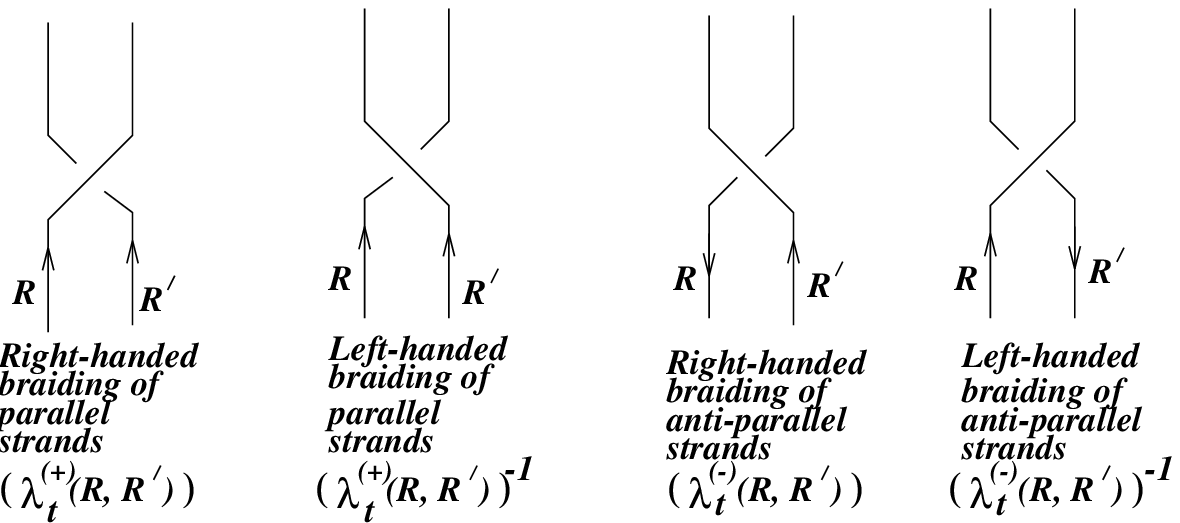}$$

As mentioned earlier, the link invariants in 
vertical framing are only regular isotopy 
invariants. That is, these invariants do not remain  unchanged  
when a writhe is smoothed out, but instead pick up a phase:
$$
\epsfbox{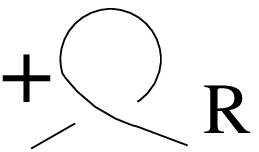} =~ 
q^{C_R}~~~\epsfbox{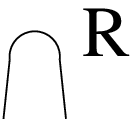}~,$$
$${\rm and}~~~\epsfbox{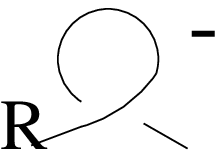}=~
 q^{-C_R}~~ ~\epsfbox{Sec5fig11.eps}~,
$$
where $\pm$ indicate the right-handedness and left-handedness of the 
writhe.

Writing the weaving pattern  ${\bf B}$
in Figure (a) above in terms of elementary braids,  the Chern-Simons
functional integral over this manifold is given by a matrix
$ {\bf B} (\{ R_i\},~ \{ \ell_i \}) $ in ${\cal{H}}_1 \, \otimes
{\cal{H}}_2$. 
To plat this braid, we consider two balls with Wilson lines as
shown in Figures (b) and (c) above. We glue  these respectively 
from above and below onto the manifold of Figure (a).  This 
yields a link in $S^3$. 

The Chern-Simons functional integral over the ball (c) is given by  
a vector in the Hilbert space associated with its $S^2$ boundary.
This vector  ~$\vert \psi (\{\ell_i \})\rangle$ ~can again be 
written in terms of a convenient basis of this Hilbert space.
Similarly, the functional integral over the ball of Figure (b)
above is a dual vector ~$\langle \psi (\{R_i \}) \vert$~ in the associated
dual Hilbert space. Gluing these two balls on to each other (along
their oppositely oriented boundaries) gives $n$
disjoint unknots carrying representations $R_1,~R_2,~....~R_n$  in $S^3$. Their
invariant is simply the product of  invariants for $n$ individual unknots,
due to the factorization property of invariants for disjoint knots. Thus 
the inner product of vectors representing the functional integrals
over manifolds (b) and (c) is given by 
\begin{equation}
\langle \psi(\{ R_i \}) \vert  \psi (\{ R_i \})\rangle~=~
\prod^n_{i~=~1} ~V_{R_i}[U] ~,
\end{equation}
where the invariant for unknot carrying representation
$R_i$ is given by $V_{R_i}[U] = dim_q ~R_i$ which is the
quantum dimension for representation $R_i$ defined in terms of
highest weight $\Lambda_{R_i}$, Weyl vector $\rho$ and
positive roots $\alpha_+$ \cite{kau5}:
\begin{equation}
{\rm dim}_q ~R_i= \prod_{\alpha_ \in \Delta_+} 
{[(\Lambda_{R_i}, \alpha_+)+ (\rho, \alpha_+)] \over [(\rho, \alpha_+)]}
= {S_{0 \Lambda_{R_i}} \over S_{0 0}} ~. \label {modst}
\end{equation}
Here the square brackets denote $q-$number defined as:
\begin{equation}
[x]~=~{(q^{\frac x 2} -q^{-\frac x 2}) \over 
(q^{\frac 1 2}- q^{-\frac 1 2})}~,
\end{equation}
with $q$ as given in eqn. (\ref {quon}). Matrix $S$ represents the 
generator of modular
transformation $\tau \rightarrow -1/\tau$ on the characters of 
associated ${\cal G}_k$ Wess-Zumino conformal
field theory. Its  form  
for a group ${\cal G}$ of rank $r$ and dimension
$d$ is given by \cite{gepn, kac, sene, fuch}:
\begin{eqnarray}
S_{\Lambda_{R_1} \Lambda_{R_2}} = (-i)^{d-r \over 2} 
\vert {L_{\omega} \over L} \vert^{-{1 \over 2}} 
\left(k + C_v \right)^{-{1 \over 2}} 
\sum_{\omega \in W} \epsilon (\omega) \exp \left( { -2 \pi i \over k+C_v} 
(\omega (\Lambda_{R_1} + \rho), \Lambda_{R_2} + \rho) \right)~,
\label{smatrix}
\end{eqnarray}
where $W$ denotes the Weyl group and its elements $\omega$
are words constructed using the generator $s_{\alpha_i}$ --
that is, $\omega = \prod_i s_{\alpha_i}$ and 
$\epsilon(\omega) = (-1)^{\ell (\omega)}$ with 
$\ell (\omega)$ as  length of the word.
The action of the Weyl generator $s_{\alpha}$ on a weight $\Lambda_R$
is
\begin{equation} 
s_{\alpha} (\Lambda_R) = \Lambda_R - 2 \alpha {(\Lambda_R, \alpha) 
\over (\alpha, \alpha)}~, 
\end{equation}
and $\vert {L_{\omega} /  L }\vert$ is the ratio of 
weight and  co-root lattices 
(equal to the determinant of
the cartan matrix for simply laced algebras).

Having determined the state corresponding to functional integrals
over three-manifolds as drawn in Figs. (b), (c) above, we 
shall now obtain the invariant for a link 
obtained by gluing 
the two balls (b) and (c) on to the manifold of Figure
(a). The link invariant is equal to    
the matrix element of  matrix ${\bf B}$ 
between these two vectors. This can be calculated by generalizing
the method of ref.(\cite {kau2,kau3}) for arbitrary semi-simple groups through
following proposition:

\vskip0.3cm

{\bf Proposition 1:}~{\it Expectation value of a Wilson
operator for an arbitrary} $n$ {\it component framed
link $[L, {\bf f}]$ with a plat
representation in terms of a  braid}
${\bf {B}}(\{ R_i \}, \{ \ell_i \})$ {\it generated as a word in 
terms of the braid generators is given by}
\begin{equation} 
V[L; {\bf f}; R_1, R_2, \ldots R_n] 
\, = \, \langle \psi(\{ R_i \}) \vert ~{\bf {B}}(\{ R_i \},
\{ \ell_i \}) ~\vert \psi (\{ \ell_i \})\rangle
\end{equation}

Thus these invariants for any arbitrary framed link can be evaluated.

\vskip0.5cm

{\bf Examples}

\vskip0.3cm

a) Unknot with framing number +1 or -1 is related to the unknot in
   zero framing as:

\begin{eqnarray}
{\epsfxsize=.5in \epsfbox{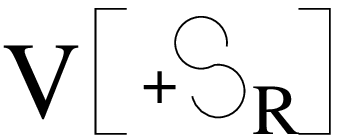}}&=&~ q^{C_R} 
~{\epsfxsize .5in \epsfbox{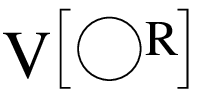}} 
=~ q^{C_R}~ dim_q R \\
{\rm and}~~~~~~~~~~~{\epsfxsize .5in \epsfbox{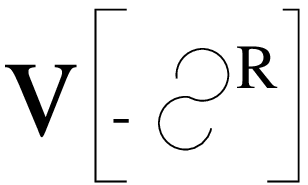}}&=&~ 
 q^{-C_R} ~{\epsfxsize .5in \epsfbox{Sec5fig13.eps}} 
=~q^{-C_R}~ dim_q R ~. 
\end{eqnarray}

b) The invariant for a Hopf link carrying representation $R_1$ and $R_2$
on the component knots in  vertical framing can be obtained in
two equivalent ways using the braiding and inverse braiding (parallel
strands):
\begin{eqnarray}
{\epsfxsize 1in \epsfbox{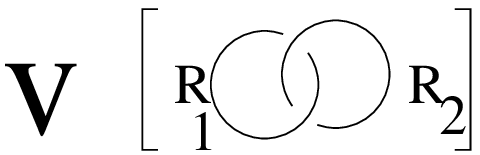}} ~&=&~ \sum_{\ell}~
N_{R_1 R_2}^{ \ell} dim_q ~\ell ~\left ( \lambda_{\ell} ^{(+)}(R_1,
~R_2) \right )^2 \nonumber\\
&=& ~q^{-C_{R_1} -C_{R_2}} ~ \sum_{\ell} 
N_{R_1 R_2}^{\ell}~ dim_q ~\ell
~~q^{C_{\ell}},  \label {hopfa}\\
{\epsfxsize 1in \epsfbox{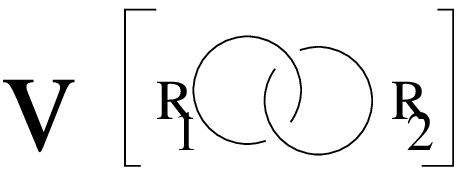}} ~&=&~ \sum_{\ell}~
N_{R_1 R_2}^{\ell} dim_q ~\ell ~
\left ( \lambda_{\ell}^{(+)} (R_1, ~R_2) \right )^{-2} \nonumber\\ 
&=& ~q^{C_{R_1} +C_{R_2}} ~ \sum_{\ell} ~ 
N_{R_1 R_2}^{\ell} dim_q ~\ell ~~q^{-C_{\ell}}, \label {hopfb}
\end{eqnarray}
where the summation over $\ell$ is over all the 
allowed representations in ${\cal G}_k$ 
conformal field theory and  coefficients 
$N_{R_1 R_2}^{\ell}$ are given by the fusion rules of the conformal theory: 
\begin{eqnarray}
N_{R_1 R_2}^{\ell}&=& 1 ~~~~~~{\rm if ~ 
 \ell \in R_1 \otimes R_2 ~~( ~fusion ~rules ~of 
~{\cal G}_k ~conformal ~field ~theory}) \nonumber\\
~&=& 0 ~~~~~~{\rm otherwise}~.
\label{fusionrules}
\end{eqnarray}
There is an explicit form for the fusion matrix in terms 
of elements of  modular matrix $S$ \cite{verl, fuch, sene}: 
\begin{equation}
N_{R_1 R_2 }^{\ell} = \sum_m S_{\Lambda_{R_2} \Lambda_m}  
\left ( {S_{\Lambda_{R_1} \Lambda_m} \over S_{0 {\Lambda_m}}} \right)
S^*_{\Lambda_m \Lambda_{\ell}} ~, \label {fusi}
\end{equation}
where 
$\left ( {S_{\Lambda_{R_1} \Lambda_m} \over S_{0 \Lambda_m}} \right)$
can be shown to be  the eigenvalues of  fusion matrix $(N_{R_1})^{\ell}_{R_2}
\equiv N_{R_1 R_2}^{\ell}$.
We can show the topological equivalence of Hopf links (\ref {hopfa},
\ref {hopfb}) by exploiting the properties of generators $S$ 
(\ref {smatrix}) and  $T$ ($\tau \rightarrow \tau + 1$) 
representing modular transformations
on the characters of  associated ${\cal G}_k$ Wess-Zumino conformal 
field theory: 
\begin{eqnarray}
S^2&=&C ~,\\
(ST)^3 &=& 1 ~,\label {ident}
\end{eqnarray} 
where $C_{\Lambda \Lambda'}=\delta_{\Lambda' {\bar \Lambda}}$~ 
is the charge conjugation matrix and
$S^*= S^{\dagger}= S^{-1}$, 
$S^*= CS = SC$; 
$S^*_{\Lambda \Lambda'} = S_{\bar {\Lambda} \Lambda'}=
S_{\Lambda \bar {\Lambda'}}$.
Modular  generator $T$ has a diagonal form\cite {gepn,fuch}:
\begin{equation}
T_{\Lambda_{R_1} \Lambda_{R_2}} = \exp({-i \pi c \over 12})~ 
q^{C_{\Lambda_{R_1}}}~ \delta_{ \Lambda_{R_1} \Lambda_{R_2}}~,
\end{equation}
where central charge of the conformal field theory is $c = {kd \over k+ C_v}$
with $d$ denoting dimension of the group ${\cal G}$. 
Eqn. (\ref {ident}) can be rewritten as 
\begin{equation}
\sum_{\Lambda_{\ell}} S_{\Lambda_m \Lambda_{\ell}} ~q^{ C_{\Lambda_\ell}}~ 
S_{\Lambda_{\ell} \Lambda_t} = \alpha ~
q^{- C_{\Lambda_m} - C_{\Lambda_t}}
~S^*_{\Lambda_m \Lambda_t}, \label{idty}
\end{equation}
where $\alpha = exp({i \pi c \over 4}).$
Using eqns. (\ref {modst}, \ref {fusi}, \ref {idty}), we have:
\begin{equation}
q^{-C_{R_1} -C_{R_2}} ~ \sum_{\ell} ~dim_q ~\ell
~~q^{C_{\ell}} ~= ~q^{C_{R_1} +C_{R_2}} ~ \sum_{\ell}
~dim_q ~\ell
~~q^{-C_{\ell}}~=~{S_{\Lambda_{R_1} \Lambda_{R_2}}\over S_{0 0}}~. 
\end{equation}
Here summation over $\ell$ runs over only those
irreducible representations in the product $R_1 \otimes   R_2$
which are allowed by fusion rules (eqn. \ref {fusionrules}). 
This confirms the equality of two expressions in eqns. (\ref {hopfa})
and (\ref {hopfb}) for the invariant for Hopf link. 

(c) \noindent Next consider  the Hopf link $H(R_1,~R_2)$ with  framing $+1$
 for each of its component knots as drawn below. The framing is represented
by a right-handed writhe in each of the knots.
\vskip0.2cm
\centerline{\epsfxsize=2in \epsfbox {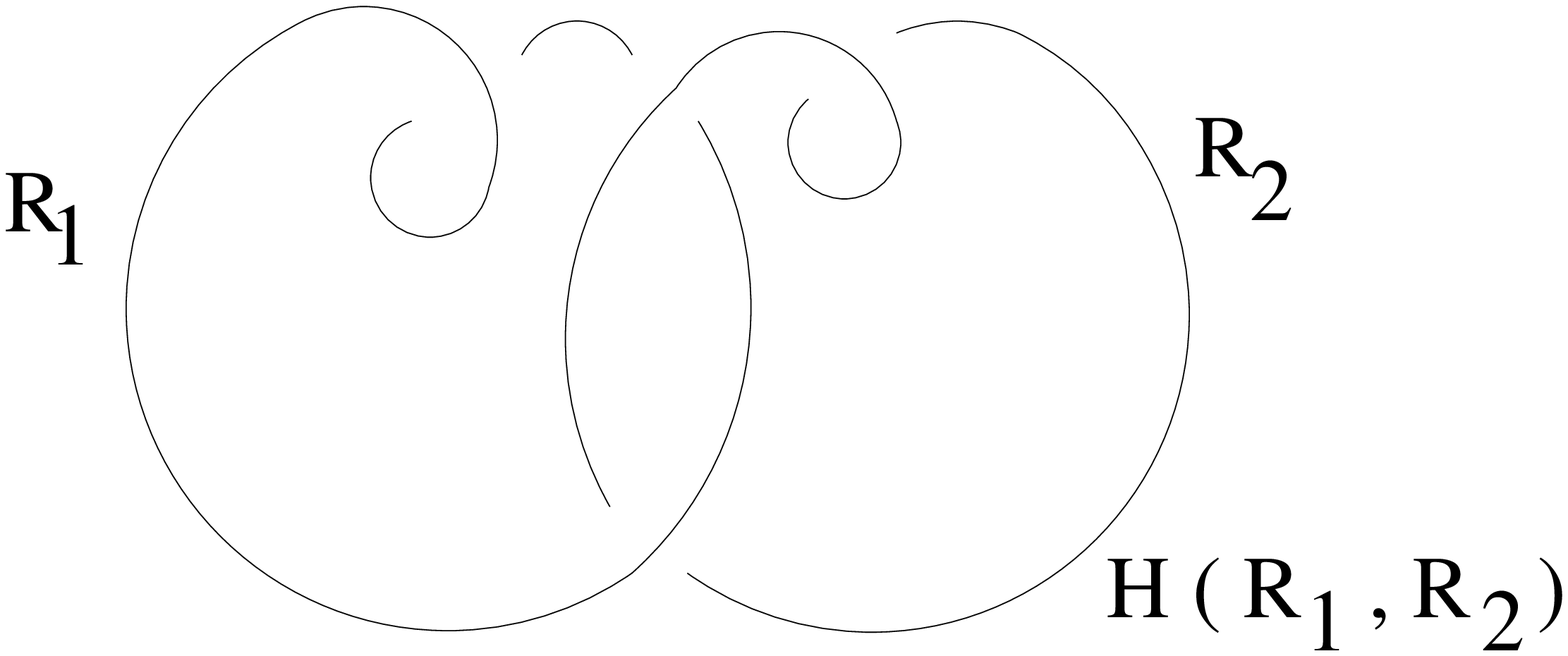}}
\vskip0.2cm

\noindent The invariant for this link is given by
\begin{equation}
V[H(R_1,~R_2)] ~=~ q^{C_{R_1} +C_{R_2}} ~~{\epsfxsize
1in  \epsfbox{Sec5fig16.eps}}
~=~ q^{C_{R_1}
+C_{R_2}}~~{S_{\Lambda_{R_1} \Lambda_{R_2}}\over S_{00}}~, \label {hopfc} 
\end{equation}
where the first factor ~$q^{C_{R_1} +C_{R_2}}$~ comes from  two writhes.

Next we shall present a discussion of how such invariants for framed
links in $S^3$  based on any arbitrary group 
can be used to construct a manifold invariant for arbitrary
three-manifolds. This will generalize a similar construction done
earlier for $SU(2)$ group \cite{kau3, kau4}.

\section{Three-manifold invariants}

We shall first recapitulate the mathematical details of how
three-manifolds are constructed by surgery on framed unoriented
links. This will subsequently be used to derive an algebraic formula
in terms of the link invariants characterizing three-manifolds
so constructed.
Starting step in this discussion  is a theorem due 
to Lickorish and Wallace \cite{wal,rol}:
\vskip0.2cm

{\bf Fundamental theorem of Lickorish and Wallace:}~{\it  Every
closed, orientable, connected three-manifold, $M^3$ can be
obtained by surgery on an unoriented framed knot or link $[L, ~f]$ 
in $S^3$.}
\vskip0.2cm
 
As pointed out earlier  framing $f$ of a link  $L$ is defined by associating 
with every component knot $K_s$ of the link an
accompanying closed curve $K_{sf}$ parallel to the knot and
winding $n(s)$ times in the right-handed direction. That is
the linking number $lk(K_s, K_{sf})$ of the component knot 
and its frame is $n(s)$. In so  
called vertical framing where the frame is thought to be
just vertically above the two dimensional projection of the
knot as shown below, we may indicate this  by 
putting $n(s)$ writhes in the  knot or even
by just simply writing the integer $n(s)$ next to the knot
as shown below: 

\centerline{\epsfbox{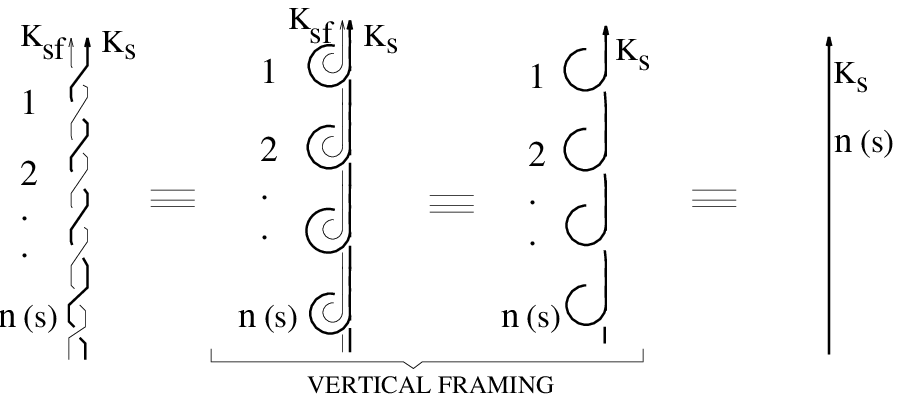}} 
\vskip0.2cm

Next the surgery on a framed link $[L, f]$  made of
component knots $K_1, K_2, ....~K_r$  with framing
$f~=~ (n(1), n(2), ....~n(r))$ in $S^3$ is performed in the following
manner. Remove a small open solid torus neighbourhood ${N_s}$
of each component knot $K_s$, disjoint from all other such
open tubular neighbourhoods associated with other component
knots. In the manifold left behind $S^3 -(N_1 \cup N_2
\cup~....~N_r)$, there are $r$ toral boundaries. On each such
boundary, consider a simple closed curve (the frame) going 
$n(s)$ times along the meridian and once along the longitude 
of  associated knot $K_s$. Now do a modular transformation on
such a toral boundary such that the framing curve bounds a
disc. Glue back the solid tori into the gaps. This yields a
new manifold $M^3$. The theorem of Lickorish and Wallace
assures us that every closed, orientable, connected
three-manifold can be constructed in this way. 

This construction of three-manifolds  is not unique: 
surgery  on more than one framed link can yield homeomorphic 
manifolds. But the rules of equivalence of framed links in
$S^3$ which yield the same three-manifold on surgery are
known. These rules are known as Kirby moves.

\vskip0.3cm

{\bf Kirby calculus on framed links in $S^3$:}
Following two elementary moves (and their inverses) generate 
Kirby calculus\cite{kir}:

\vskip0.3cm

{\it Move I}.~ For a number of unlinked strands belonging
to the component knots $K_s$ with framing $n(s)$ going
through an unknotted circle $C$ with framing $+1$, the
circle can be removed after making a
complete clockwise (left-handed) twist from below in the disc enclosed 
by  circle $C$: 

\centerline{\epsfbox{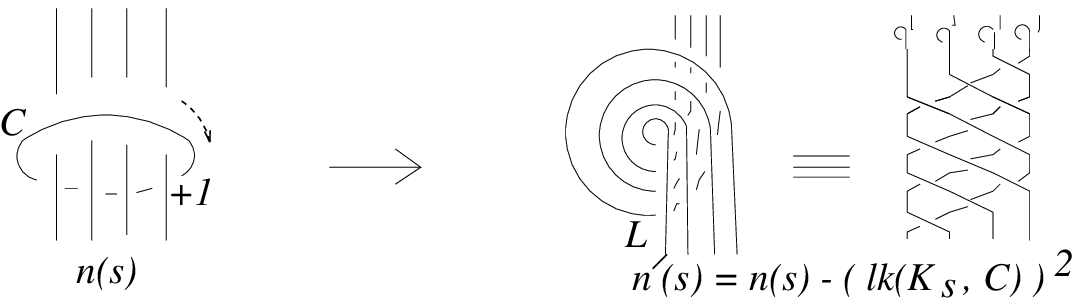}} 
\vskip0.2cm

\noindent  In the process, in addition to
introducing new crossings, the framing of the various
resultant component knots, $K'_s$ to which the affected  strands
belong, change from $n(s)$ to $n'(s)~=~n(s)- \left (
lk(K_s, C) \right )^2$.

\vskip0.3cm

{\it Move II}. ~Drop a disjoint unknotted circle $C$ with
framing $-1$ without any change in  rest of the
link:

\centerline{\epsfbox{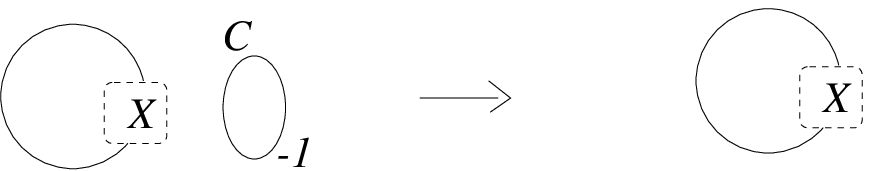}} 

\vskip0.2cm

\par Two Kirby moves (I) and (II) and their inverses
generate the conjugate moves\cite{lic}:
\vskip0.2cm

{\it Move ${\bar I}$}. ~Here a circle $C$ with framing $-1$
enclosing a number strands can be removed after
making a complete anti-clockwise (right-handed) twist from below in
the disc bounded by  curve $C$:

\centerline{\epsfbox{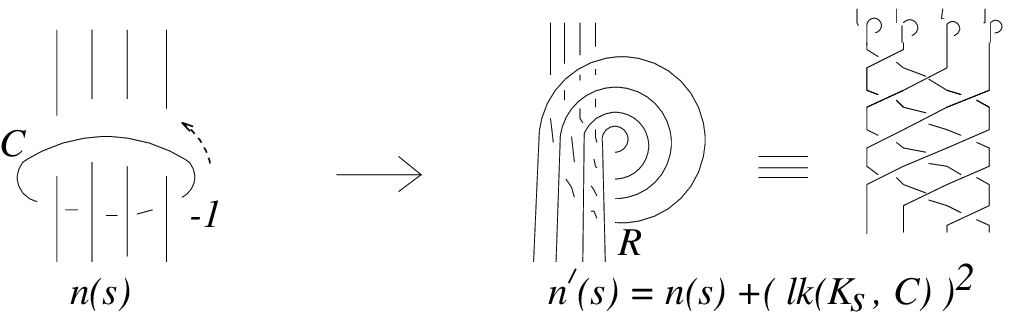}} 
\vskip0.2cm

\noindent Again, this changes the framing of the
resultant knots $K'_s$ to which the enclosed strands
belong from $n(s)$ to $n'(s)~=~n(s) + \left (
lk(K_s, C) \right )^2$.
\vskip0.2cm

{\it Move ${{\bar I}{\bar I}}$}.~ A disjoint unknotted circle $C$
with framing $+1$ can be dropped without affecting
rest of the link:

\centerline{\epsfbox{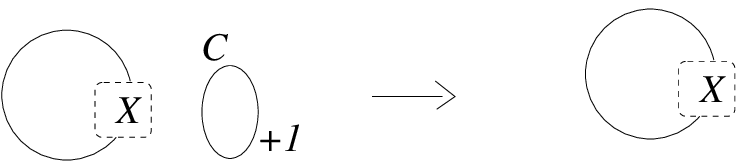}} 
\vskip0.2cm 

\par  Thus Lickorish-Wallace theorem  and equivalence of
surgery under Kirby moves  reduces the theory of closed,
orientable, connected three-manifolds to the theory of
framed unoriented links via a one-to-one correspondence:
$$
\left( \matrix {Framed~ unoriented~ links~ in~ S^3~ modulo \cr
 equivalence~under~ Kirby ~moves} \right) ~\leftrightarrow
~
\left( \matrix
{Closed, ~orientable,~ connected~ three-\cr manifolds~ 
modulo~ homeomorphisms} \right)
$$
\noindent This consequently allows us to characterize 
three-manifolds by the invariants of  associated 
unoriented framed knots and links obtained from the
Chern-Simons theory in $S^3$. This can be done by
constructing an appropriate combination of the invariants
of framed links which is unchanged under Kirby moves and 
which does not see orientations of the framed link:
$$
\left( \matrix {Combination~ of~ framed~ ~link ~invariants \cr
  which~ do~ not~ change ~under~ Kirby ~moves} \right) ~=~~
\left( \matrix
{Invariants ~ of ~associated \cr three-manifold~
} \right)
$$
Using the framed link invariants presented in 
the previous section,
we shall now construct such a three-manifold invariant which 
is preserved under Kirby moves.
The immediate step in this direction  is to construct a 
combination of these link invariants
which would be unchanged under Kirby move I:

\vskip0.2cm
\centerline{\epsfbox{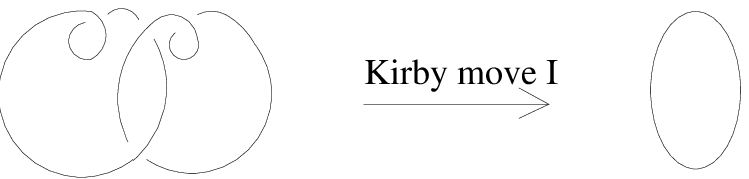}}
\vskip0.2cm

\noindent In order to achieve this, we have to first 
solve the  following equation  for $\mu_{R_2}$ and
$\beta$ relating the invariants $V[H(R_1, R_2)] $ and
$V[U(R_1)]$ for these two links respectively: 
\begin{equation}
\sum_{R_2} ~\mu_{R_2}~V[ H(R_1,R_2)] ~=~ \beta ~V[U(R_1)] \label {idta} 
~,
\end{equation}
\noindent where summation $R_2$ is over
all the representations 
(highest weight $\Lambda_{R_2}$) with projection along the longest
root $\theta$ as
$(\Lambda_{R_2}, \theta) \leq k$. These are all the integrable
representations of ${\cal G}_k$ conformal field
theory. Rewriting the framed link invariants in terms
of modular transformation matrix $S$ (\ref {hopfc}, \ref {modst})
and comparing with the identity (\ref {idty}), we deduce the
following solution:
\begin{equation}
\mu_{R_2} ~=~ S_{0 \Lambda_{R_2}}~, ~~~~~\beta =\a \equiv e^{\pi ic/4}~.
\end{equation}

Next we will consider the following two links 
$H(X;~ R_1,~R_2)$  and $U(X;~R_1)$: 
\vskip0.2cm
\centerline{\epsfbox{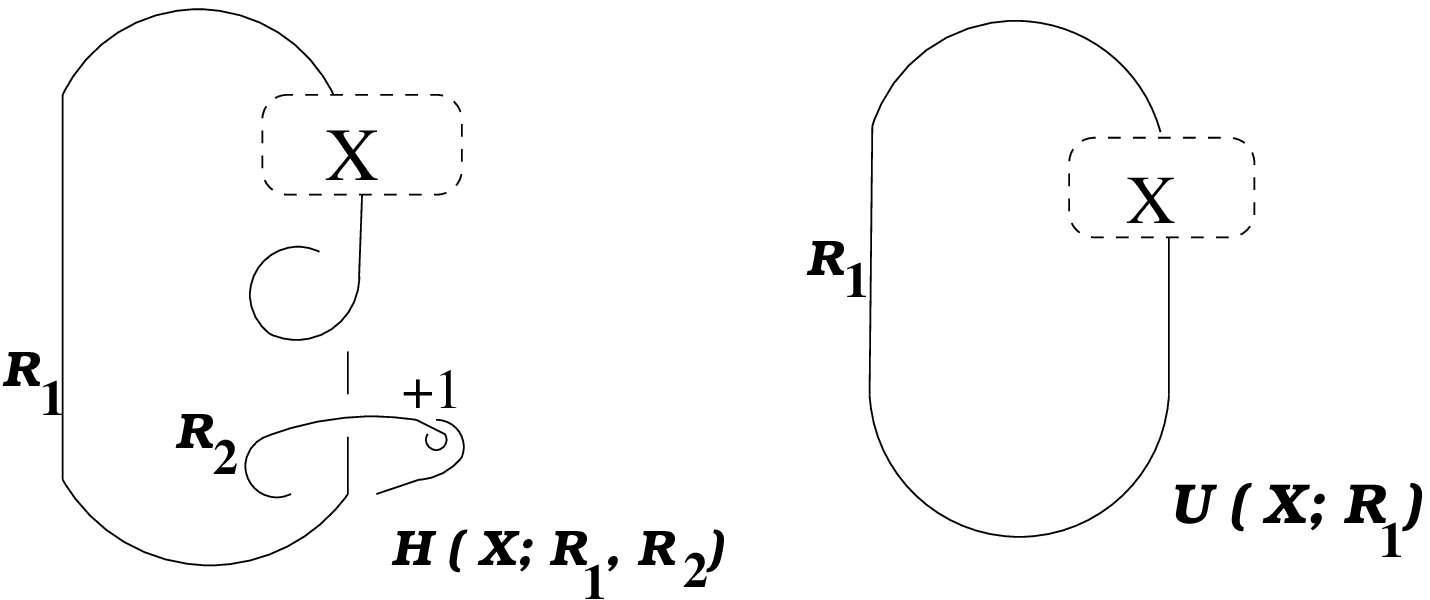}}
\vskip0.2cm

\noindent where  $X$ as an arbitrary entanglement inside the
box. The link $H(X;~R_1,~R_2)$ is  connected sum of the link $U(X;~ R_1)$
and a framed Hopf link $H(R_1,~R_2)$. Factorization properies of invariants of
such a connected sum of links yields:
\begin{equation}
dim_q~ R_1~ ~V[H(X;~R_1,~R_2)]~=~V[U(X;~R_1)]~~V[H(R_1,~R_2)]~.
\end{equation}
\noindent Using eqn. (\ref{idta}), this further implies (note 
$dim_q ~R_1$ is the invariant $V[U(R_1)]$ for unknot
in representation $R_1$ and with zero framing):
\begin{equation}
\sum_{R_2}~ \mu_{R_2}~V[H(X;~R_1,~R_2)] ~=~ \alpha~V[U(X;~R_1)]~ \label{res1}.
\end{equation}
\noindent We can generalize this  relation
for the following links $H(X;~R_1,~R_2,~....~R_n;~R)$ and $U(X;~R_1,~R_2,~..
..R_n)$,
\vskip0.2cm
\centerline{\epsfxsize 3in \epsfbox{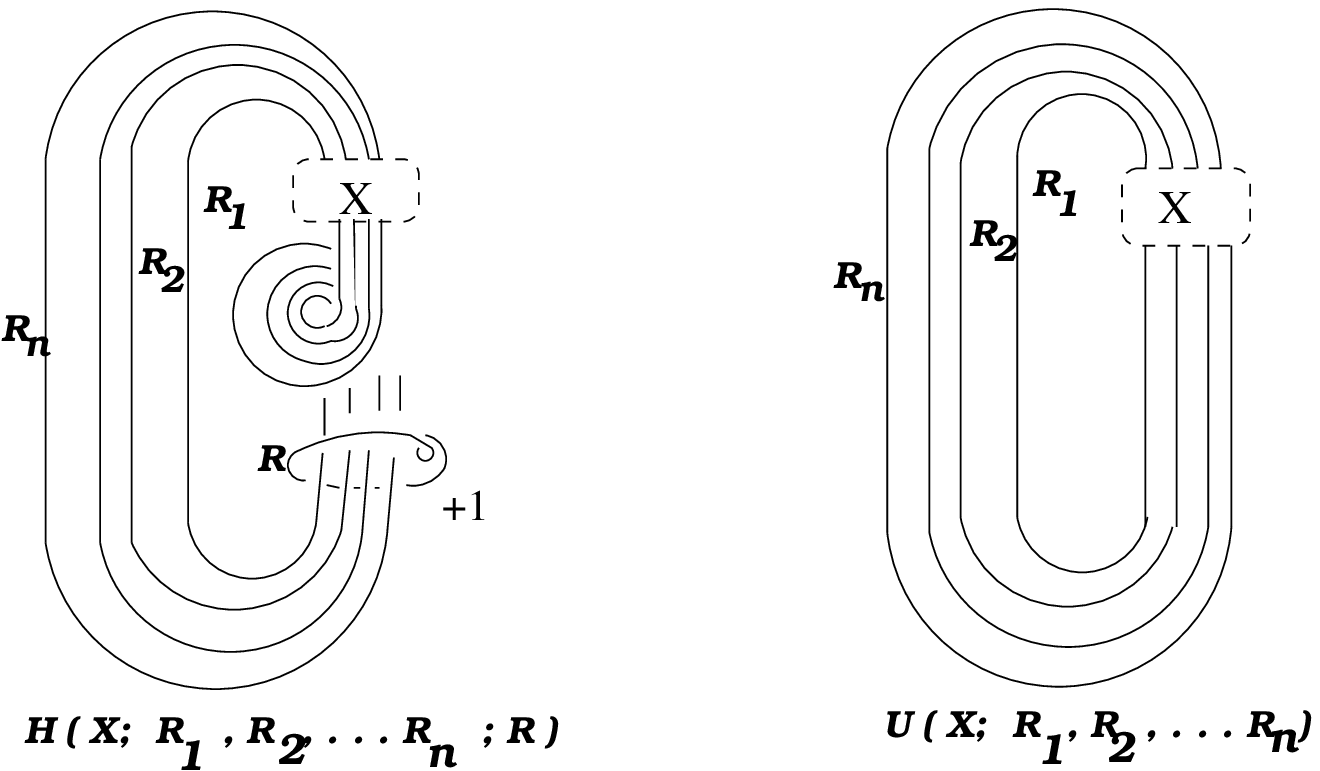}}
\vskip0.2cm
\noindent as

\noindent
{\bf Proposition 2}: {\it The invariants for these two links are related as:}
\begin{equation}
\sum_R~ \mu_R~V[H(X;~R_1,~R_2,...~R_n; ~R)] ~=~
\alpha~V[U(X;~R_1,~R_2, ....~R_n)]~.\label{kir1}
\end{equation}
Thus this proposition provides for equivalence of the two links under Kirby
move I up to a phase factor $\alpha$ on the right-hand side. Lets us now
outline the proof of this proposition.

{\bf Proof}:
In the following figure, we  have redrawn the link  
$H(X; R_1, R_2, \ldots R_n;~ R)$ in $S^3$ 
by gluing four three-manifold:
two three-balls (each with $S^2$ boundary) and two three-manifolds with
two $S^2$ boundaries each. The various boundaries  have been glued together 
along the dotted lines as indicated. 
This allows us to use the Proposition 1 above to evaluate the invariant
for this link.
\vskip0.2cm
\centerline{\epsfxsize 2in \epsfbox{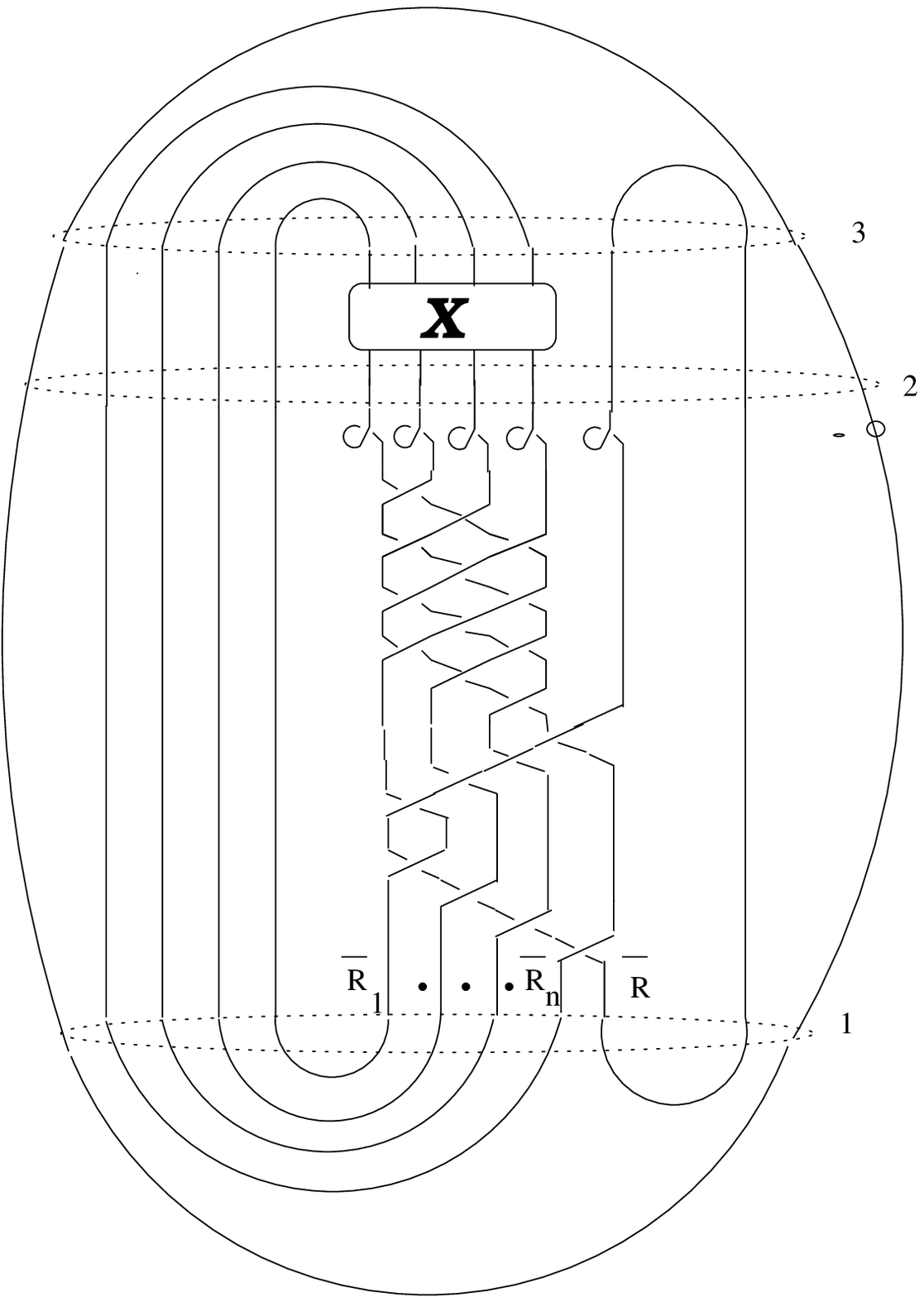}}
\vskip0.2cm
Instead of plat diagram described in Section 2, 
we have the  closure of a braid here. The
states corresponding to Chern-Simons functional integral
over two three-balls  with  $S^2$ boundaries  `1' and  `3' will be
represented by the vector  $\vert \psi(\{R_i\}, R) \rangle$ 
 (where $i \in [1,n]$) and its dual. 
The matrix element corresponding to the braiding inside the three-manifold  
with two $S^2$ boundaries `1' and `2'  can be explicitly computed  
in a convenient  basis. We shall work with a  basis represented by the following
conformal block of  associated Wess-Zumino theory: 
\vskip0.1cm
\centerline{\epsfxsize 2in \epsfbox{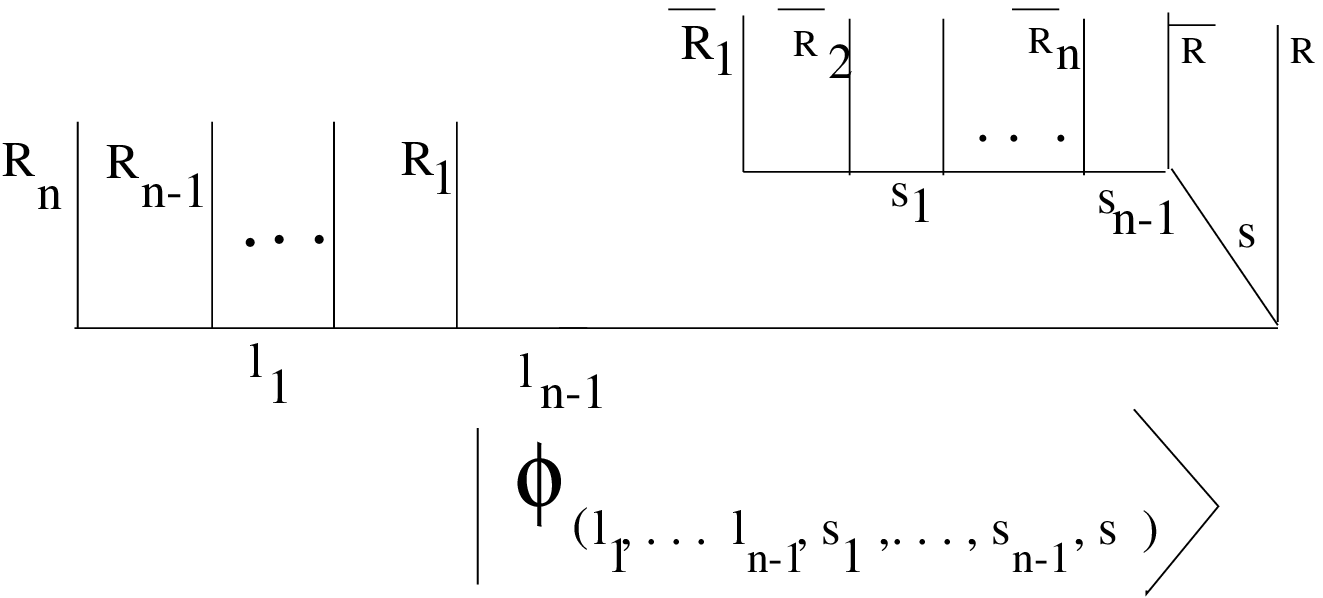}}~
\vskip0.1cm
\noindent
where $l_1 = R_n \otimes R_{n-1}, \ldots l_{n-1} = l_{n-2} \otimes R_1$
and $s_1 = R_1 \otimes R_2, \ldots s_{n-1} = s_{n-2} \otimes R_n, s=
l_{n-1} \otimes R$.
In this basis the matrix corresponding to the three-manifold with
boundaries marked `1' and `2' turns out to be:
\begin{equation}
\nu_1 = \sum_{l_1, \ldots l_{n-1}, s_1, \ldots s_{n-1}, s} q^{C_s} 
~~\vert \phi_{(l_1, \ldots l_{n-1}, s_1, \ldots s_{n-1}, s)}^{(2)} 
\rangle \langle \phi_{(l_1, \ldots l_{n-1}, s_1, \ldots s_{n-1}, s)}^{(1)}
\vert \label {dualt} 
\end{equation}
where the superscripts $(1)$  and $(2)$ inside the
basis states refer to  two $S^2$ boundaries containing the 
three-manifold.
This result involves properties of the braiding and duality 
matrices which we present for $n = 2$  in the Appendix. These can readily
be generalized to arbitrary $n$.

The matrix $\nu_2 (X)$  representing the other three-manifold  containing
entanglement $X$ between  two $S^2$ boundaries indicated by dotted 
lines `$2$' and `$3$' in the figure above can similarly be evaluated.
In addition, we also need to write down the states $\vert\psi^{(1)}\rangle$
and $\langle \psi^{(3)} \vert $  corresponding to the two
three-balls  with boundaries indicated by dotted lines `$1$'
and `$3$'. All these  in the above  basis can be written as :
\begin{eqnarray}
\nu_2 (X)&=& \sum_{\{l_i\}, s, \{s_i\}, \{s^{\prime}_i\}} 
X (\{s_i\}, ~ \{s^{\prime}_i\})~~ 
\vert \phi_{(l_1, \ldots l_{n-1}, s_1, \ldots s_{n-1}, s)}^{(3)} 
\rangle \langle \phi_{(l_1, \ldots l_{n-1}, s^{\prime}_1, \ldots 
s^{\prime}_{n-1}, s)}^{(2)}\vert~\\
\vert \psi ^{(1)} \rangle &=& \sum_ {l_1, \ldots l_{n-1},  , s} \sqrt {dim_q ~s} 
~~\vert \phi_{(l_1, \ldots l_{n-1}, l_1, \ldots l_{n-1}, s)}^{(1)}
\rangle\\ 
\langle \psi ^{(3)} \vert &=& \sum_ {l_1, \ldots l_{n-1}, s} \sqrt {dim_q ~s} 
~~\langle \phi_{(l_1, \ldots l_{n-1}, l_1, \ldots l_{n-1}, s)}^{(3)}
\vert
\end{eqnarray} 
Gluing these four  three-manifolds  along the oppositely oriented
$S^2$ boundaries, we get the link $H(X;R_1, R_2, \ldots R_n; R)$ 
whose invariant can now be written as:
\begin{equation}
V[H(X;R_1, R_2 \ldots R_n; R)] = \sum_{\{l_i\}}
 X(\{l_i \},~ \{l_i \})\left[ \sum_{s \in l_{n-1} \otimes
R} (dim_q~ s )~~q^{C_s}\right]~. \label {ghopf}
\end{equation}
Clearly the term in square bracket is the Hopf link
invariant $H(l_{n-1}, R)$ of eqn. (\ref {hopfc}).
Similarly, we can compute the link invariant for $U(X; R_1, R_2 \ldots
R_n)$ as
\begin{equation}
V[U(X;R_1, R_2 \ldots R_n)] = \sum_{\{l_i\}}
 (dim_q ~l_{n-1}) ~~X(\{l_i \},~ \{l_i \})~. \label {gunot}
\end{equation}
Now using eqn.(\ref {idta}),  it is easy to prove Proposition 2:
\begin{eqnarray}
\sum_R \mu_R~ V[H(X; R_1, R_2 \ldots R_n; R)]&=&\sum_{\{l_i\}} 
X(\{l_i\}, ~ \{l_i \}) \sum_R \mu_R~
V[H(l_{n-1},R)]  \\
~&=& \alpha ~V[U(X; R_1, R_2 \ldots R_n)]~.
\end{eqnarray} 

This completes our discussion of Kirby move I. For Kirby move II,
we note that for  a link containing a disjoint unknot with framing $-1$, 
we have:
\begin{equation}
\sum_{\ell}~ \mu_{\ell}~\epsfbox{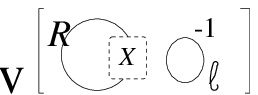}  ~=~
\alpha^* ~\epsfbox{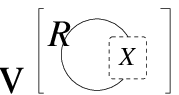} \label{kir2}
\end{equation}
\noindent This follows readily due to the exact
factorizations of invariants of  disjoint links into those
of  individual links and use of the identity:
~~$\sum_{\ell}~ S_{0 \ell} ~q^{-C_{\ell}}~S_{\ell 0} ~=~
e^{-\pi ic/4}~ S_{0 0}$.

Clearly the Eqns. (\ref{kir1}) and (\ref {kir2})
respectively correspond to the two generators of  Kirby
calculus.  Presence of phases $\alpha$ in these equations reflects
change of three-framing of the  associated manifold under  Kirby moves. 
Three-manifolds
constructed by surgery on links equivalent under Kirby moves
though  topologically same,  yet may differ in terms of their
three-framing. The dependence on three-framing ($\alpha$ factors in
above equations representing Kirby moves)
can be rotated away by use of a nice property of the linking matrix
under Kirby moves.
For a framed link $[L,~f]$ whose component knots
$K_1,~K_2,~....~K_r$ have framings (self-linking numbers) as 
$n_1,~n_2,~....~n_r$
respectively, the linking matrix is defined as
$$
W[L,~f]~=~\left ( \matrix { n_1 &lk(K_1,K_2) &lk(K_1,K_3)
&.....&lk(K_1,K_r) \cr
lk(K_2,K_1) & n_2& lk(K_2,K_3)&.....&lk(K_2,K_r) \cr
..&..&n_3&.....&.. \cr
..&..&..&.....&..\cr
lk(K_r,K_1)&..&..&.....&n_r} \right )
$$
\noindent where $lk(K_i,K_j)$ is the linking number of knots $K_i$
and $K_j$. The signature of linking matrix is given by
$$
\sigma [L,~f]~=~({\rm no. ~of ~+ve ~eigenvalues~of}~ W)-(
{\rm no.~of~-ve~eigenvalues~of}~W)
$$
\noindent Then this signature for the framed link $[L,~f]$
and those for the links $[L',~f']$ obtained by
transformation under two elementary
generators of  Kirby calculus are related in a simple
fashion:
\begin{eqnarray}
{\rm ~Kirby~ move~ I}:~~~~~~~\sigma [L,~f]~&=&~ \sigma ~[L',~f']+1~; \label {signa}
\nonumber \\
{\rm ~Kirby~move~II} :~~~~~~~\sigma [L,~f] ~&=&~\sigma ~[L',~f']-1~ \label {signb}.
\end{eqnarray}
Notice, though the sign of linking numbers $lk (K_i, K_j)$ 
for distinct knots does depend on the relative orientations of knots 
$K_i$ and $K_j$, the signature of linking matrix does not
depend on the relative orientations of component knots.

Now collecting the properties of  framed link
(\ref {kir1}, \ref {kir2}) and the signature
of linking matrix under the Kirby moves (\ref {signa}), 
we may state our main  result:

\vskip0.2cm
{\bf Proposition 3:}~{\it For a framed link $[L,~f]$ with
component knots, $K_1,~K_2,~....~K_r$ and their framings
respectively as $n_1,~n_2,~....~n_r$, the quantity}
\begin{equation}
{\hat F[L,~f]}~=~ \alpha^{-\sigma[L,~f]}~
\sum_{\{R_i\}}~~\mu_{R_1}~\mu_{R_2}~....~\mu_{R_r}~ V[L;~n_1,~n_2,~...~n_r;~
R_1,~R_2,~....~R_r] \label {inva}
\end{equation}
\noindent {\it constructed from invariants $V$ (in vertical framing)
of the 
framed link, is an invariant of the associated three-manifold
obtained by surgery on that link.}

Notice individual link invariants $V[L;n_1, n_2, \ldots n_r; R_1, \ldots
R_r]$ do in general depend on the relative orientations of 
component knots. Reversal of orientation on a particular knot changes
the group representation living on it to its conjugate. Since all 
representations are summed for each component knot, the resultant
combination (\ref {inva}) is an invariant of unoriented link.

The combination ${\hat F[L,~f]}$ of link invariants so constructed is exactly
unchanged under Kirby calculus. Because of the factor depending on signature
of linking matrix in front, there are no extra factors of $\alpha$ generated by
Kirby moves.

This generalizes a similar proposition obtained earlier\cite{kau3,kau4} 
for $SU(2)$ theory to any arbitrary semi-simple gauge group.

\vskip.5cm

\noindent
{\bf Explicit  examples:}~~
Now let us give the value of this
invariant for some simple three-manifolds.

1) The surgery descriptions of manifolds $S^3$, $S^2 \times S^1$,
$RP^3$ and Lens spaces of the type ${\cal L}(p, \pm 1)$
are given by an unknot with framing $+1,~0$ , $+2$ and $\pm p$
respectively. As indicated above the knot invariant for an unknot with zero  
framing carrying  representation $R$ is $ dim_q ~R = S_{\Lambda_R 0}/S_{0 0}$.
Thus the invariant for $S^3$ is:
$$
{\hat F}[S^3]\ = \ \alpha^{-1} \sum_{\Lambda_R} ~S_{0 \Lambda_R}~q^{C_{\Lambda_R}}~ 
{\frac {S_{\Lambda_R 0}} {S_{0 0}}} ~, 
$$
\noindent where the factor~ $q^{C_{\Lambda_R}}$ ~comes from framing $+1$ (one right-handed
writhe).  Use of identity  (\ref{idty}), ~$\sum_{\Lambda_R}~ S_{0 \Lambda_R}
~q^{C_{\Lambda_R}}~S_{\Lambda_R 0 } = \alpha~ S_{0 0}$ immediately yields the invariant
simply to be: 

\begin{equation}
{\hat F}[S^3]\ = \  1
\end{equation}

\noindent Next for three-manifold $S^2 \times S^1$, we have
\begin{equation}
{\hat F}[S^2 \times S^1] \ =\ \sum_{\Lambda_R}~ S_{\Lambda_R 0}~ 
{\frac {S_{\Lambda_R 0}} {S_{0 0}}}
\ =\  {1 \over S_{0 0}}~,\\
\end{equation}
where orthogonality of the ~$S$-matrix has been used.
\noindent Finally, for $RP^3$ and more generally  Lens spaces ${\cal L} (p, \pm
1)$, we have 
\begin{eqnarray}
{\hat F}[RP^3]&=&
\alpha^{-1} \sum_R{S_{0 \Lambda_R} ~q^{2C_{\Lambda_R}}~ S_{ \Lambda_R 0}
\over S_{00}}~,\\
{\hat F}[{\cal L} ( p, \pm 1)]&=&
\alpha^{\mp 1} \sum_R{S_{0 \Lambda_R} ~q^{\pm p C_{\Lambda_R}}~ S_{ \Lambda_R 0}
\over S_{00}}~.
\end{eqnarray}

2) A more general example is the whole class of Lens spaces
${\cal L}(p,~q)$; above manifolds are special
cases of this class. These are obtained\cite{rol}
by surgery on a framed link made of
successively linked unknots with framing given by integers
$a_1,~a_2,~......~a_n$:

\centerline{~\epsfbox{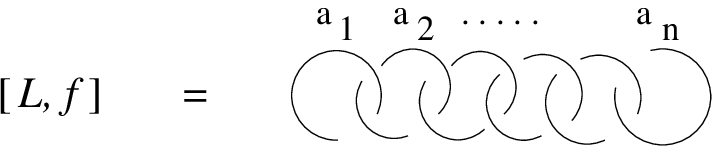}}
\vskip0.2cm

\noindent
where these framing integers provide a continued fraction
representation for the ratio of two integers $p,~q$:
$$
{p \over q} ~=~ a_n -{1 \over{a_{n-1} -{1 \over{.......~a_3- {1
\over {a_2 -{1 \over a_1}}}}}}}~.
$$
\noindent The invariant for these manifolds can readily be evaluated.
The relevant link $[L,~f]$ above is just a connected sum of framed ~$n - 1$~
Hopf links so that its link invariant is  obtained by the factorization
property of invariants for such a connected sum. Placing representations
$R_1$, $R_2$, ..... $R_n$ on the component knots, this link invariant is:

\begin{equation}
V[L, ~f;~ R_1, R_2, ....~R_n] ~=~ {\frac {q^{\sum_{i=1}^n {a_i C_{R_i}}}~~
{\prod_{i=1}^{n-1} S_{\Lambda_{R_i} \Lambda_{R_{i+1}}}}}  { S_{0 0}~~
{\prod_{i=2}^{n-1} ~~ S_{{\Lambda_{{R_i}}} 0}}}}, \nonumber
\end{equation}
where the factor ~$q^{\sum_{i=1}^n {a_i C_{R_i}}}$~ is due to the
framing $f = (a_1, ~a_2, ....~a_n)$  of  knots. This finally yields a
simple formula for the three-manifold invariant:
\begin{equation}
{\hat F}[{\cal L}(p,~q)]~=~
\alpha^{-\sigma[L,~f]}~{\alpha}^{(\sum a_i )/3}~
~{(S~M^{(p,~q)} )_{00} \over S_{00}}~,
\end{equation}
\noindent where  matrix $M^{(p,~q)}$ is given in terms of the modular
matrices $S$ and $T$:
\begin{equation}
M^{(p,~q)}~=~ T^{a_n} S~ T^{a_{n-1}} S~ ......~T^{a_2}
S~ T^{a_1} S~.
\end{equation}
\noindent The corresponding expression for the partition function for
$SU(2)$ Chern-Simons field theory in Lens spaces  has also been obtained earlier
in refs. \cite{jeff}.

3) Another example we take up is the Poincare manifold
$P^3$ (also known as dodecahedral space or Dehn's homology sphere). 
It is given \cite{rol} by surgery on a right-handed trefoil knot with 
framing $+1$:
\vskip0.2cm
\centerline{\epsfxsize 1in \epsfbox{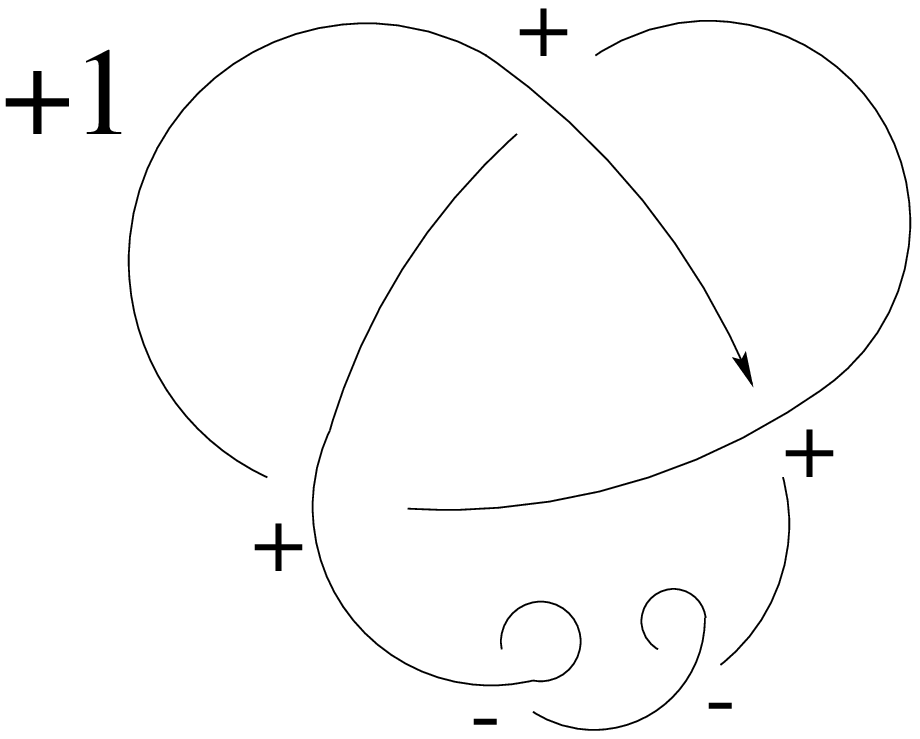}}
\vskip0.2cm
\noindent Notice, each right-handed crossing of the trefoil
introduces $+1$ linking number between the knot and its vertical
framing, and each of the two left-handed writhes contributes $-1$ so
that the total frame number  is $+1$.
The knot invariant for a right-handed trefoil (in vertical framing with no extra
writhes) carrying representation $\ell$ is:
$$
V[T;~\ell]= \sum_{R} N^{R}_{\ell \ell} 
dim_q ~R~ (-)^{6 \epsilon_{\ell} -3\epsilon_R}~ 
q^{-3C_{\ell} + {3 \over 2} C_{R}} 
$$
Using this trefoil invariant and the Proposition 3, 
the three-manifold invariant  for Poincare manifold turns out to be:
\begin{equation}
{\hat F}[P^3]~=~{\alpha}^{-1}~~\sum_{m,\ell,R}
{(-)^{6 \epsilon_{\ell} - 3\epsilon_R}  {S_{0 \Lambda_{\ell}}~S_{0 \Lambda_R}~
S_{ \Lambda_{\ell} \Lambda_m}~ S_{ \Lambda_{\ell} \Lambda_m}~
~S^*_{ \Lambda_R \Lambda_m}~q^{-5C_{\ell} +{3 \over 2}C_R}}
\over {S_{00}~S_{0 \Lambda_m}}}~.
\end{equation}

4) Similarly, surgery on a right-handed trefoil $T$ with framing number
$-1$ (that is,  with four left-handed writhes in vertical framing) yields 
another homology three-sphere (with fundamental group presented by
$\alpha, \beta:$ $(\alpha \beta)^2 = \alpha^3 = \beta^7$ ). Its invariant is
\begin{equation}
{\hat F}[T,-1]= \alpha ~\sum_{m,\ell,R}
{(-)^{6 \epsilon_{\ell} -3 \epsilon_R}  
{S_{0 \Lambda_{\ell}}~S_{0 \Lambda_R}~
S_{ \Lambda_{\ell} \Lambda_m}~
S_{ \Lambda_{\ell} \Lambda_m}~
S^*_{ \Lambda_R \Lambda_m}~q^{-7C_{\ell} +{3 \over 2}C_R}}
\over {S_{00}~S_{0 \Lambda_m}}}~.
\end{equation}

5) The surgery on a right-handed trefoil with framing number $+3$
yields a coset manifold $S^3 / T^*$ where $T^*$ is
the binary tetrahedral group generated by two
different $ 2 \pi/3$ rotations  $\alpha, \beta $ about two different
vertices of a tetrahedron with
 $\alpha^3 = \beta^3= (\alpha \beta)^2=1.$ The three-manifold
invariant for this coset manifold is 
\begin{equation}
{\hat F}[{S^3 / T^*}] = \alpha^{-1}
~\sum_{m,\ell,R}
{(-)^{6 \epsilon_{\ell} - 3\epsilon_R}  {S_{0 \Lambda_{\ell}}~S_{0 \Lambda_R}~
S_{ \Lambda_{\ell} \Lambda_m}~ S_{ \Lambda_{\ell} \Lambda_m}~
~S^*_{ \Lambda_R \Lambda_m}~q^{-3C_{\ell} +{3 \over 2}C_R}}
\over {S_{00}~S_{0 \Lambda_m}}}~.
\end{equation}

\section{Conclusions}

We have presented here a construction of a class of three-manifold invariants,
one each for any arbitrary semi-simple gauge group. 
The construction exploits the 
one-to-one correspondence between framed unoriented links in $S^3$ modulo 
equivalence under Kirby moves to closed orientable connected three-manifolds 
modulo homeomorphisms. Three-manifolds are characterized by an appropriate
combination of invariants of the associated links. This combinations of
link invariants is obtained from Chern-Simons theory in $S^3$ and  
is unchanged by Kirby moves. The construction
is a direct generalization of that developed for an $SU(2)$ Chern-Simons theory
earlier\cite{kau3, kau4}. The manifold invariant obtained from $SU(2)$
theory has been shown to be related to  partition function of Chern-Simons
theory on that manifold \cite {ram2}.
The generalized three-manifold invariants $\hat F(M)$ constructed here
are also related to the partition function $Z(M)$ of Chern-Simons theory
by an overall normalization:
\begin{equation}
\hat F(M) = {Z(M) \over  S_{00}}~.
\end{equation}
Thus, given a surgery presentation of a three-manifold, this
provides a simple method of computing 
partition function of Chern-Simons field theory based on an arbitrary
gauge group in that three-manifold.

\vskip 1cm

\noindent 
{\bf Acknowledgements:}
P.R would like to thank TIFR for hospitality where
part of this work got done. The work of P.R is supported
by a CSIR Grant.

\vskip1cm

\appendix {\bf Appendix:}

In this appendix, we shall use the properties of braiding and 
duality matrices  to derive  the result (\ref {dualt}).
for the case $n = 2$. 

A convenient basis state for $n=2$ is provided  by  the conformal
block of the Wess-Zumino theory pictorially represented as:
\vskip0.2cm
\centerline{\epsfbox{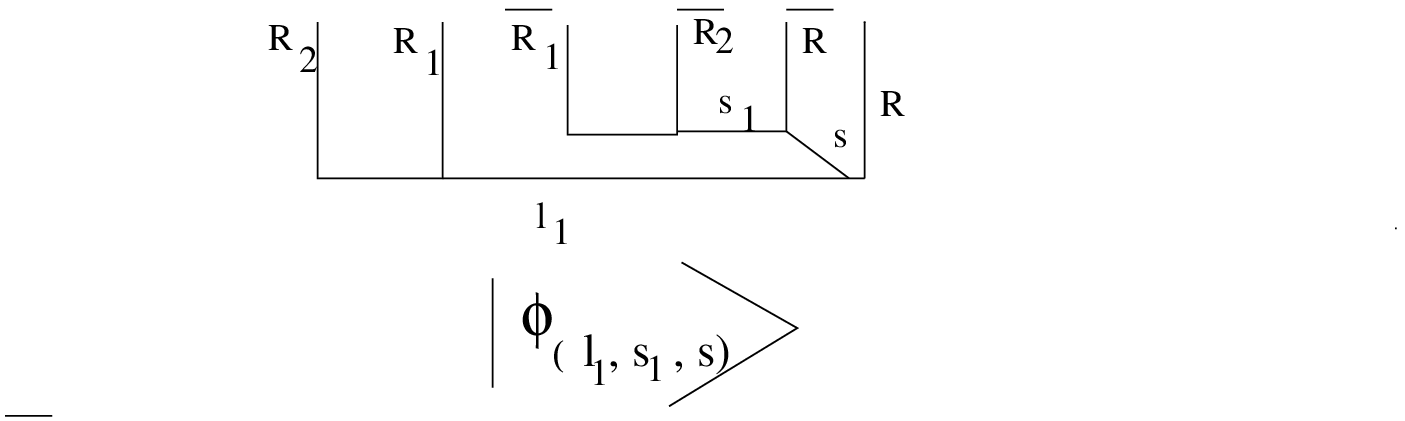}}
\vskip0.2cm

The matrix $\nu_1 $ for $n=2$ corresponds to a functional integral over a
three-manifold  with two $S^2$ boundaries, each with six punctures
carrying representations $R_2$, $R_1$, $R_1$, $R_2$, $R$ and $R$.

Between these two boundaries we have braiding $b_4~ b_3^2~ b_4~ b_3^2$
and a  $+1$ writhe on each of the three strands carrying
representations $\bar {R_1}$, $\bar {R_2}$, $\bar R$. Using the explicit representation for 
braid generators in terms of their eigenvalues and duality matrices
$a_{sp} \left[\matrix{R_1& R_2 \cr R_3& R_4}\right]
$, we obtain $\nu_1$ to be: 
\begin{eqnarray}
\nu_1 &=&q^{C_{R_1}+C_{R_2} +C_R}
~b_4~b_3^2~b_4~b_3^2 ~~| \phi_{l_1, s_1,s}^{(1)} \rangle 
\langle \phi_{l_1, s_1,s}^{(2)} | \nonumber\\
&=&q^{C_{R_1}+C_{R_2} +C_R}
 ~(\lambda_{s_1} (R_1, R_2))^2 \sum_{p_1 s_1', r, s_1''} a_{s_1 p} 
\left[\matrix{\bar {R_1}& \bar {R_2} \cr \bar R& s}\right] 
\lambda_p(R_2, R)\nonumber\\
~&~& a_{s_1' p} \left[ \matrix{\bar {R_1}& \bar R \cr \bar {R_2}& s} \right]
 (\lambda_{s_1'}(R_1,R))^2
~a_{s_1' r} \left[\matrix{\bar {R_1} & \bar R \cr \bar {R_2}& s}\right] 
\lambda_r (R, R_2) ~a_{s_1'' r} 
\left[\matrix{\bar {R_1}& \bar {R_2}\cr \bar R& s}\right] \nonumber\\
~&~& |\phi_{l_1, s_1'',s}^{(1)} \rangle
\langle \phi_{l_1, s_1,l}^{(2)} | 
\end{eqnarray}
Using the following property of the duality matrices,
\begin{eqnarray} 
\sum_l (-1)^{\epsilon_l}~ a_{p l}\left[\matrix{R_1 R_4 \cr
R_3 R_2}\right]~q^{C_l \over 2} 
~a_{p' l}\left[\matrix{R_1 R_3 \cr
R_4 R_2}\right]
&=& (-1)^{\epsilon_{R_1}+ \epsilon_{R_2} +
\epsilon_{R_3}+ \epsilon_{R_4}-\epsilon_p -\epsilon_{p'}}\nonumber\\
&~&a_{p p'}\left[\matrix{R_3 R_2 \cr R_4 R_1}\right] ~q^{-C_p- C_{p'} + C_{R_1}
+ C_{R_2}+C_{R_3}+C_{R_4} \over 2}
\end{eqnarray}
and the orthogonality relation,
the above equation can be
simplified to give 
\begin{equation}
\nu_1 = 
\sum_{l_1,s_1,s} q^{C_s}~|\phi_{l_1,s_1,s}^{(2)} \rangle
\langle \phi_{l_1,s_1,s}^{(1)}| 
\end{equation} 
Generalization of this result for arbitrary $n$ is straight forward.


\begin{thebibliography} {}
\bibitem{ati} M. Atiyah: {\it The Geometry and Physics of Knots},
Cambridge University Press (1989).
\bibitem{wit}E. Witten: Quantum field theory and Jones
polynomials, Commun. Math. Phys. {\bf 121} (1989) 351-399.
\bibitem{kau1} R.K. Kaul and T.R. Govindarajan: Three dimensional
Chern-Simons theory as
a theory of knots and links, Nucl. Phys. {\bf B380} (1992)
293-333;  Three dimensional
Chern-Simons theory as a
theory of knots and links II: multicoloured links, Nucl.
Phys. {\bf B393} (1993) 392-412.
\\
P. Ramadevi, T.R. Govindarajan and R.K. Kaul: Three
dimensional Chern-Simons
theory as a theory of knots and links III: an arbitrary
compact semi-simple
group,  Nucl. Phys.{\bf B402} (1993) 548-566. \\
R.K. Kaul: Knot invariants from quantum field theories,
in {\it Modern QFT II 1994:} 271-284, Editors. S.R. Das, G. Mandal,
S. Mukhi and S.R. Wadia, World Scientific 1995.
\bibitem{kau2}R.K. Kaul: Complete solution of $SU(2)$ Chern-Simons
theory, hep-th/9212129; \\
Chern-Simons theory, coloured-oriented braids and link
invariants, Commun. Math. Phys. {\bf 162} (1994) 289-319. 
\bibitem{lab1} J.M.F. Labastida, P.M. Llatas and A.V. Ramallo: 
Knot operators in Chern-Simons gauge theory, 
Nucl. Phys. {\bf 348} (1991) 651; \\
J.M.F. Labastida and M. Marino: The HOMFLY polynomial
for torus links from Chern-Simons gauge theory, Int. Journ. 
Mod. Phys. {\bf A10} (1995) 1045.
\bibitem{kau3} R.K. Kaul: Chern-Simons theory, knot invariants,
vertex models and three-manifold invariants, hep-th/9804122, 
in {\it Frontiers of Field Theory, Quantum Gravity and 
Strings (Volume 227 in Horizons in World Physics)}, eds. R.K. Kaul 
et al, NOVA Science Publishers, New York (1999). 
\bibitem{lab2} J.M.F. Labastida: Chern-Simons gauge theory: ten years after,
hep-th/9905057.
\bibitem{kau4} R.K. Kaul: Topological Quantum Field theories --
A meeting ground for physicists and mathematicians, hep-th/9907119,
 in {\it Quantum Field Theory: A 20th Century Profile}, ~
ed. A.N. Mitra, Hindustan Book Agency and Indian National Science
Academy (New Delhi)  (2000) pp 211-232.
\bibitem{lic} W.B.R. Lickorish: $3$-manifolds and the
Temperley Lieb algebra, Math. Ann. {\bf 290} (1991)
657-670; 
 Three-manifold invariants from combinatorics of
Jones polynomial, Pac. J. Math. {\bf 149} (1991)
337-347.                                                                
\bibitem{ram1} P. Ramadevi, T.R. Govindarajan, R.K. Kaul: Representations of
Composite Braids and Invariants for mutants knots and links in
Chern-Simons field theories, Mod. Phys. Lett. {\bf A10} (1995) 1635-1658.
\bibitem{ram2} P. Ramadevi and Swatee Naik: Computation of Lickorish's
three-manifold invariants using Chern-Simons theory, hep-th/9901061,
Commun. Math. Phys. {\bf 209} (2000) 29-49.
\bibitem{res}N.Y. Reshetikhin and V. Turaev: Invariants of
three manifolds via link polynomials and quantum groups,
Invent. Math. {\bf 103} (1991) 547-597.
\bibitem{bir} J.S. Birman: {\it Braids, Links and Mapping
Class groups}, Annals of Mathematics Studies Princeton Univ.
Press (1975). 
\bibitem{kau5}R.K. Kaul: The representations of Temperley-Lieb-Jones algebras, 
 Nucl. Phys. {\bf B417} (1994) 267-285.
\bibitem{gepn} D. Gepner and E. Witten: String Theory on 
Group manifold, Nucl. Phys. {\bf B278} (1986) 493.
\bibitem{kac} V.G. Kac and D. H. Peterson: Infinite-dimensional 
Lie Algebras, Theta functions and Modular 
forms, Adv. Math. {\bf 53} (1984) 125.
\bibitem{sene} P. Di Francesco, P. Mathieu, D. Senechal: Conformal
Field Theory - Graduate Texts in Contemporary Physics, eds. J. L. Birman,
J. W. Lynn, M.P. Silverman, H. E. Stanley, Mikhail Voloshin; 
Springer-verlag (1997).
\bibitem{fuch} J. Fuchs: Affine Lie Algebras and Quantum Groups,
Cambridge Monographs on Mathematical Physics, Cambridge Univ. Press.
\bibitem{verl} E. Verlinde: Fusion rules and modular 
transformations in 2D conformal field theory, 
Nucl. Phys. {\bf B300} [FS22] (1988) 360. 
\bibitem{wal} A.D. Wallace: Modification and
cobounding manifolds, Can. J. Math. {\bf 12} (1960)
503. 
\bibitem{rol} D. Rolfsen: {\it Knots and links}, Publish
or Perish, Berkeley (1976).
\bibitem{kir} R. Kirby and P. Melvin:  On the $3$-manifold invariants of
Reshetikhin-Turaev for $sl(2,C)$, Invent. Math. {\bf 105}
(1991) 473-545;\\ 
B.R. Lickorish: A representation of orientable
combinatorial $3$-manifold, Ann. of Math. {\bf 76}
(1962) 531.
\bibitem{jeff} L.C. Jeffrey: Chern-Simons-Witten invariants of Lens
spaces and torus bundles and the semi-classical approximation,
Commun. Math. Phys. {\bf 147} (1992) 563-604; \\
S. Kalyana Rama and S. Sen: 3D manifolds, graph invariants 
and Chern-Simons theory, Mod. Phys. Lett. {\bf A7} (1992) 2065 and
Comments on Witten invariants of three-manifolds for $SU(2)$ and
$Z_m$: Mod. Phys. Lett. {\bf A8} (1993) 2285. 


\end{thebibliography}
\end{document}